\documentclass[aps,prl,twocolumn,amsmath,amssymb,superscriptaddress]{revtex4}

\newcommand{\bra}[1]{\langle#1|}
\newcommand{\ket}[1]{|#1\rangle}

\usepackage[pdftex]{graphicx}
\usepackage{mathrsfs}
\usepackage{times}
\usepackage{color}

\begin{document}

\bibliographystyle{apsrev}

\title{The information capacity of a single photon}

\author{Peter P. Rohde}
\email[]{dr.rohde@gmail.com}
\homepage{http://www.peterrohde.org}
\affiliation{Centre for Engineered Quantum Systems, Department of Physics and Astronomy, Macquarie University, Sydney NSW 2113, Australia}

\author{Joseph F. Fitzsimons}
\affiliation{Centre for Quantum Technologies, National University of Singapore,
Block S15, 3 Science Drive 2, Singapore 117543}

\author{Alexei Gilchrist}
\affiliation{Centre for Engineered Quantum Systems, Department of Physics and Astronomy, Macquarie University, Sydney NSW 2113, 
Australia}

\date{\today}

\frenchspacing


\begin{abstract}
Quantum states of light are the obvious choice for communicating quantum information. To date, encoding information into the polarisation states of single photons has been widely used as these states form an natural closed two state qubit. However, photons are able to encode much more -- in principle infinite -- information via the continuous spatio-temporal degrees of freedom. Here we consider the information capacity of an optical quantum channel, such as an optical fibre, where a spectrally encoded single photon is the means of communication. We use the Holevo bound to calculate an upper bound on the channel capacity, and relate this to the spectral encoding basis and the spectral properties of the channel. Further, we derive analytic bounds on the capacity of such channels, and in the case of a symmetric two-state encoding calculate the exact capacity of the corresponding channel.
\end{abstract}

\maketitle


\emph{Introduction} --- Single photons are an ideal candidate for efficiently communicating both quantum and classical information \cite{bib:NielsenChuang00}. Unlike many other quantum systems, photons are inherently `flying', making them ideal for quantum communication tasks, including quantum key distribution and distributed quantum computation. In optical quantum information processing \cite{bib:KLM01, bib:KokLovett11}  it is common to encode a qubit into the polarisation of a single photon. That is, a qubit is defined as $\alpha \ket{H} + \beta \ket{V}$, or a classical bit can be communicated by choosing $\ket{H}$ or $\ket{V}$. Alternately, an encoding could be performed in the photon-number or quadrature bases. These cases have been studied extensively by previous authors \cite{bib:Pierce81, bib:Yamamoto86, bib:Drummond94, bib:Wolf07, PhysRevA.54.1869, PhysRevA.63.032312, PhysRevLett.92.027902, PhysRevA.72.042330}. For example, the Fock basis, $\{\ket{n}\}$, could be employed to encode an alphabet with a number of letters limited only by energy constraints. While the alphabet may in principle be arbitrarily large, once loss is introduced or physically realistic encoding procedures and photo-detectors are employed, which introduces mixing in the photon number degree of freedom, the information capacity is limited.

In this Letter we approach photonic information capacity from an entirely different perspective. We fix the number of photons at \mbox{$n=1$}, and encode information into its spectral degree of freedom \cite{bib:MlburnCoherent08}. Since the spectral degree of freedom is continuous, in principle infinite information could be transmitted by a single photon encoded in this basis. However, subject to realistic channel, detector and photon engineering constraints, the communicable information is reduced. We examine the information capacity of a single photon via encoding in the spectral domain and derive bounds on the channel capacity using such an encoded photon under realistic assumptions about the communications channel and photo-detector. We relate the channel capacity to the spectral response of the channel and photo-detector, and the choice of spectral encoding basis.


\emph{The spectral structure of photons} --- A photon can be expressed as a superposition of different spectral components, allowing an $N$-level qudit to be encoded, where $N$ can in principle be arbitrarily large. To perform such encoding we choose a set of spectral functions $\{\psi_i(\omega)\}$, where $\omega$ is frequency relative to a central carrier frequency. Ideally we would like these functions to form an orthonormal basis, \mbox{$\int_{-\infty}^\infty \psi_i(\omega)\psi_j(\omega)^* \,\mathrm{d}\omega = \delta_{i,j}$}, such that they can always be perfectly distinguished with an appropriate measurement device. In reality, however, orthogonality might only be approximate. We define photonic mode operators \cite{bib:RohdeMauererSilberhorn07} which create photons with a well defined spectral distribution function $\psi(\omega)$, \mbox{$A_{\psi(\omega)}^\dag = \int_{-\infty}^\infty \psi(\omega) a^\dag(\omega)\,\mathrm{d}\omega$}, where $a^\dag(\omega)$ is the single frequency photonic creation operator in spatial mode $a$. We will employ the shorthand $A_{\psi_i(\omega)}^\dag\ket{0}\equiv\ket{i}$, where $\ket{0}$ is the vacuum state. Then a spectral basis state may be expressed as \mbox{$\rho_i = A_{\psi_i(\omega)}^\dag \ket{0}\bra{0} A_{\psi_i(\omega)} = \ket{i}\bra{i}$}.

In principle, any basis $\{\psi_i(\omega)\}$ could be chosen, such as frequency or temporal delta functions, wavelet families, Hermite polynomials, or any other set of functions satisfying orthonormality. However, photon engineering \cite{bib:Branning00, bib:Grice01, doi:10.1080/0950034021000011455, PhysRevA.66.065403, bib:URen03, bib:URen05, bib:Torres05, PhysRevA.73.063802, bib:MlburnCoherent08} is an emerging field and not all states can be readily prepared on-demand with sufficient fidelity.


\emph{Information capacity of a single photon} --- Let Alice encode classical information by choosing $i$ in the range $(1,N)$. Thus, a single photon sends a letter from an $N$ letter alphabet. Next we propagate the spectrally encoded state through a channel (such as an optical fibre), which has a frequency-dependent transmission function $\eta(\omega)$. That is, the channel has probability $\eta(\omega)^2$ of propagating a photon of frequency $\omega$, otherwise it is absorbed by the channel. Here we will assume that the channel is Markovian, as this assumption holds true for most physical mechanisms inducing photon loss, and hence we are free to model the channel as a frequency dependent beamsplitter \cite{bib:RohdeRalph06b}, where the reflected component is traced out. The spectral response of a photo-detector after the channel can be merged with the spectral response of the channel, \mbox{$\eta(\omega)=\eta_\mathrm{channel}(\omega)\eta_\mathrm{detector}(\omega)$}. 


\begin{figure}[!htb]
\includegraphics[width=0.8\columnwidth]{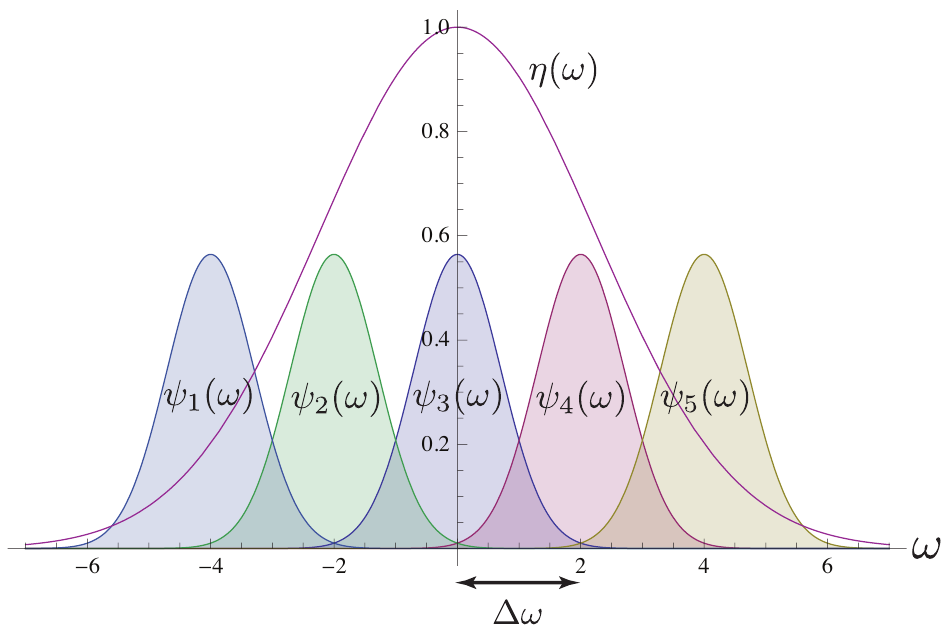}
\caption{(Colour online) Example choice of encoding basis $\{\psi_i(\omega)\}$, where the basis states are not perfectly orthogonal, and an example channel spectral response function $\eta(\omega)$. The functions $\{\psi_i(\omega)\}$ are identical Gaussians of different means, separated in frequency by multiples of $\Delta \omega$. By letting the Gaussians have zero variance, one could achieve orthonormality and encode a qudit with an arbitrary number of levels. However, this would be physically unrealistic. In this example we have centred $\omega$ around $0$, which is unphysical. Thus, $\omega$ should be interpreted as being relative to some central carrier frequency. All the integral overlaps defining the channel capacity are invariant under uniform translations in $\omega$, so the choice of centre frequency does not affect the results. Other examples of encoding bases are time-bin encoding, which would yield the same results in the conjugate domain, or orbital angular momentum encoding.
} \label{fig:sample_basis}
\end{figure}

When a spectrally encoded basis state passes through this channel, the output state is (after tracing over the environment), \mbox{$\rho_i' = A_{\eta(\omega)\psi_i(\omega)}^\dag \ket{0}\bra{0} A_{\eta(\omega)\psi_i(\omega)} + \epsilon_i \ket{0}\bra{0}$}, where \mbox{$\epsilon_i = \int_{-\infty}^\infty | \sqrt{1-\eta(\omega)^2} \psi_i(\omega) |^2 \,\mathrm{d}\omega$}. Thus, after the channel we have a mixture of the vacuum state (corresponding to the absorbed component), and a single photon with spectral distribution function modulated by the spectral response of the channel. Note that $A_{\eta(\omega)\psi_i(\omega)}^\dag \ket{0}$ is in general not normalised.

We wish to establish how much information Alice is able to communicate to Bob using her spectrally encoded single photon state across the channel. Because the spectral response of the channel modulates the spectral basis states, in general the optimal choice of measurement basis for Bob will not be the same as Alice's encoding basis. To place an upper bound on the mutual information between Alice and Bob, we calculate the Holevo bound \cite{bib:Holevo73, bib:NielsenChuang00}, which bounds the mutual information under \emph{any} choice of measurement basis for Bob. Formally, the mutual information between Alice and Bob is bounded by the Holevo quantity as \mbox{$H(A:B) \leq S(\rho') - \sum_i p_i S(\rho_i')$}, where $\rho' = \sum_i p_i \rho_i'$ and $p_i$ is the a priori probability that basis state $i$ will be transmitted. We emphasise that the Holevo bound is merely an upper bound on the mutual information, which, in general, cannot always be saturated.

The mixture observed by Bob is \mbox{$\rho' = \sum_i p_i (A_{\eta(\omega)\psi_i(\omega)}^\dag \ket{0}\bra{0} A_{\eta(\omega)\psi_i(\omega)} + \epsilon_i \ket{0}\bra{0})$}. The terms in this mixture have been modulated and are in general no longer orthonormal. We will re-express the output state in some orthonormal basis $\{\phi_k(\omega)\}$,
\begin{eqnarray}
\rho' &=& \sum_{i=1}^N p_i \sum_{j,j'} (\lambda_j^{(i)} {\lambda_{j'}^{(i)}}^* A_{\phi_j(\omega)}^\dag \ket{0}\bra{0} A_{\phi_{j'}(\omega)} + \epsilon_i \ket{0}\bra{0}) \nonumber \\
&=& \sum_{j,j'} Y_{j,j'} \ket{j}\bra{j'} + \epsilon \ket{0}\bra{0},
\end{eqnarray}
where \mbox{$\lambda_j^{(i)} = \int_{-\infty}^\infty \phi_j(\omega) \psi_i(\omega)^* \eta(\omega)\,\mathrm{d}\omega$}, \mbox{$Y_{j,j'} = \sum_{i=1}^N p_i \lambda_j^{(i)} {\lambda_{j'}^{(i)}}^*$}, and \mbox{$\epsilon = \sum_{i=1}^N p_i \epsilon_i$}. Then it can be calculated that
\begin{eqnarray}
S(\rho_i') &=& -\epsilon_i\, \mathrm{log}_2 \epsilon_i - (1-\epsilon_i) \mathrm{log}_2 (1-\epsilon_i), \nonumber\\
S(\rho') &=& - \epsilon\, \mathrm{log}_2 \epsilon - \sum_{j=1}^N Y_j' \mathrm{log}_2 Y_j',
\end{eqnarray}
where $Y_j'$ is the $j$th eigenvalue of $Y$. Thus, the Holevo bound is
\begin{eqnarray} \label{eq:holevo}
&&H(A:B) \leq - \epsilon\, \mathrm{log}_2 \epsilon - \sum_j Y_j' \mathrm{log}_2 Y_j' \nonumber \\
&&+ \sum_{i=1}^N p_i [\epsilon_i\, \mathrm{log}_2 \epsilon_i + (1-\epsilon_i) \mathrm{log}_2 (1-\epsilon_i)].
\end{eqnarray}
The Holevo bound is maximised by optimising over $p_i$, which may be prohibitive for large $N$. 


\emph{Classical channel capacity} --- In photonic quantum computation \cite{bib:KLM01, bib:KokLovett11} it is common to accommodate for lossy channels via post-selection. That is, we discard events where the wrong number of photons are measured due to photon loss. In the case of a communications channel, both post-selected and non-post-selected scenarios are useful, and we will consider these two scenarios separately as they are suited to inherently different situations. First, note that if a photon is sent, detection of ``no photon'' actually contains information about the encoded state, and it is therefore in general sub-optimal to post-select out such events. Specifically, the loss of a photon gives us information that the encoded basis state was more likely to be in the region where $\eta(\omega)$ is low. For example, consider Fig. \ref{fig:sample_basis}. In this example, if no photon is detected it is more likely that Alice's encoded letter was $\psi_1(\omega)$ or $\psi_5(\omega)$ than $\psi_3(\omega)$. Thus, photon loss communicates information from Alice to Bob, which would be discarded if post-selection were introduced into the protocol. 

In the case of a constant bit-rate communications channel, where photons are being transmitted at predictable regular intervals, this observation leads us to conclude that it is best not to post-select and instead interpret photon loss as a legitimate signal.  In this case, the classical channel capacity is simply \mbox{$\mathcal{C}\leq H(A:B)$}. On the other hand, with a variable bit-rate channel, there is no way of knowing whether measurement of the vacuum state corresponded to photon loss or simply lack of a transmission. In this case post-selection is necessary and \mbox{$\mathcal{C}\leq(1-\epsilon)S(\rho_\mathrm{PS}')$}, where $\rho_\mathrm{PS}'$ is $\rho'$ post-selected on there being a photon.


\emph{Numeric results} --- We now calculate an upper bound on the capacity of the channel in a specific subset of encodings, as quantified by the Holevo bound, before going on to derive general bounds on the capacity of such channels later in this Letter. We consider a spectral basis comprised of fixed-width, displaced Gaussians, $\psi_j(\omega) \propto e^{-(\omega-(j-1)\Delta\omega)^2/4}$, each offset by $\Delta \omega$ from the next, as shown in Fig. \ref{fig:sample_basis} \footnote{Different photon source technologies will produce photons with various spectral structures, such as Lorentzians. But we use Gaussians as an illustrative example. In this example, $\Delta\omega$ can be regarded as a measure of photon distinguishability: \mbox{$\Delta\omega=0$} for indistinguishable photons, and \mbox{$\Delta\omega\gg 1$} for distinguishable photons. Obviously, since the energy, \mbox{$E=\hbar \omega$}, must be finite, there are fundamental physical limitations on how many basis states may be employed. However we will not consider energy constraints in our analysis, and assume that all basis states are physically realisable. Gaussians are technically unphysical since they have non-zero amplitude for zero and negative frequencies. However, centred around a realistic carrier frequency these components are negligible.}. Optimising \mbox{$H(A:B)$} over $p_i$ for large $N$ is prohibitive, so in our numeric analyses we will make the simplifying assumption that Alice is employing a uniform encoding, \mbox{$p_i=1/N$} \footnote{Note that a uniform encoding is in general not optimal. However, in the setting where Alice knows nothing about the channel, making the assumption that the channel has a flat response, and she additionally believes she is preparing frequency delta functions, although her photon engineering is imperfect and is in fact yielding Gaussians with non-zero overlap, then employing a uniform encoding is a justifiable choice for Alice.}.

\begin{figure*}[!htb]
\includegraphics[width=0.62\columnwidth]{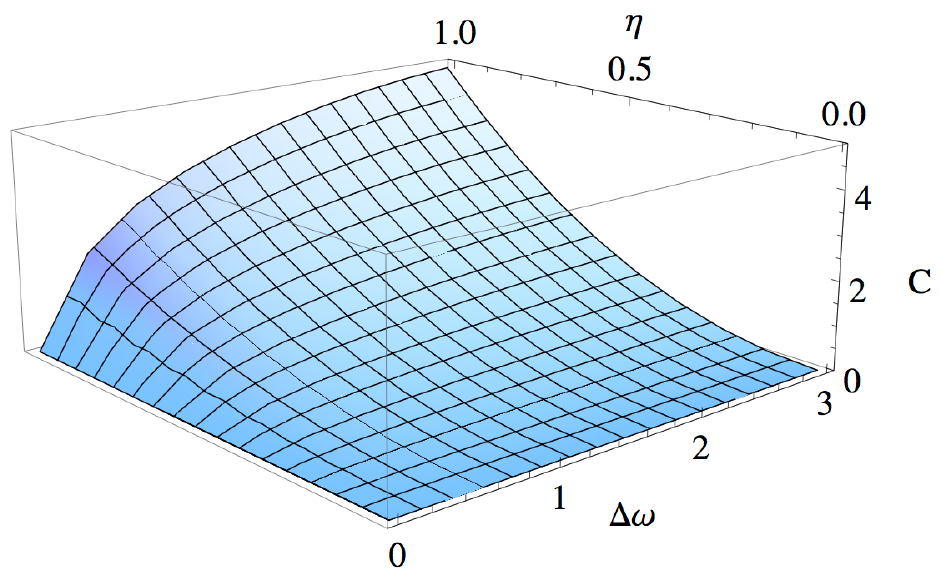} 
\includegraphics[width=0.71\columnwidth]{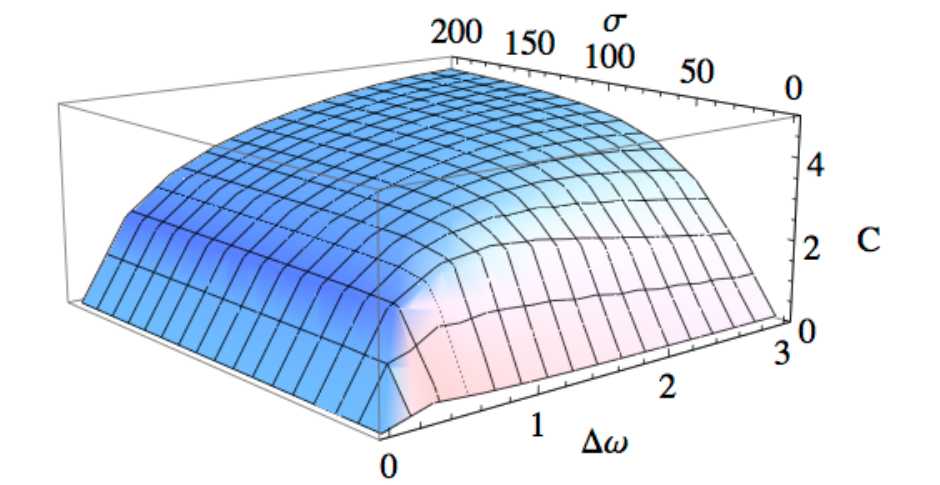}
\includegraphics[width=0.71\columnwidth]{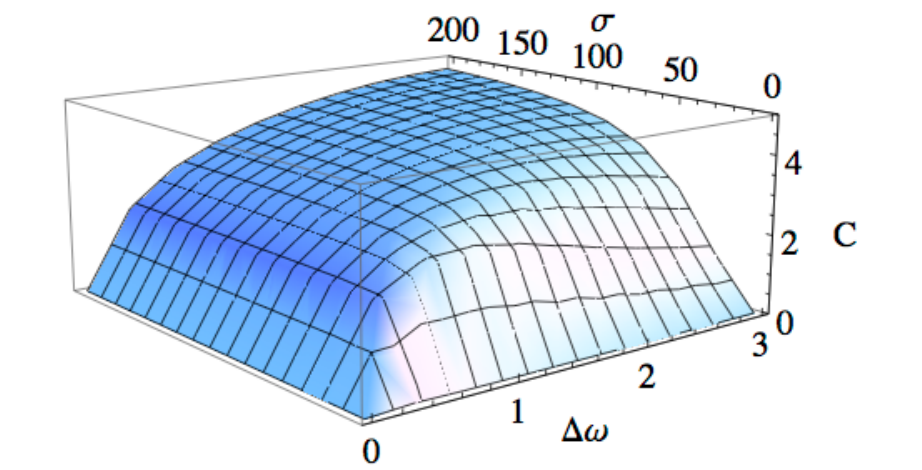}
\caption{(Colour online) Upper bound on the classical channel capacity (bits per transmitted photon) using Gaussian wave-packets, displaced by multiples of $\Delta \omega$, where Alice employs uniform encoding, \mbox{$p_i=1/N$}. (left) Flat spectral response channel (efficiency $\eta$), with or without post-selection. (middle) Gaussian spectral response channel (bandwidth $\sigma$), without post-selection. (right) Gaussian spectral response channel, with post-selection. We encode across $N=32$ basis states, a maximum of $\mathrm{log}_2 N=5$ bits of information. In the limit of \mbox{$\Delta\omega \gg 1$} (distinguishable photons), and \mbox{$\sigma \gg 1$} or \mbox{$\eta=1$} (a perfect channel), the upper bound on $\mathcal{C}$ asymptotes to the maximum achievable $\mathrm{log}_2 N$. In this limit $\mathcal{C}$ saturates the bound.} \label{fig:results}
\end{figure*}

We consider two situations. First, when the channel spectral response function is a constant, $\eta(\omega)=\eta$, and second, when the channel spectral response function is Gaussian \footnote{The motivation for employing a Gaussian spectral response is that optical channels, such as fibres, are typically tailored to a certain frequency, and their transmissive properties tail away for light away from the desired frequency.} with standard deviation $\sigma$, \mbox{$\eta(\omega)=e^{-(\omega/\sigma)^2/4}$}. The upper bound on $\mathcal{C}$ is plotted in Fig. \ref{fig:results}, where we encode across $N=32$ spectral basis states, for a maximum of \mbox{$\mathrm{log}_2 N=5$} bits of information.

For a flat spectral response function we observe a monotonic increase in $\mathcal{C}$ as both the wave-packet separation (i.e. photon distinguishability) and channel efficiency increase. In the limit of large $\Delta \omega$ and \mbox{$\eta=1$}, we observe the upper bound is the maximum achievable \mbox{$\mathcal{C}\leq \mathrm{log}_2 N$} bits. It is obvious that $\mathcal{C}$ must increase monotonically with $\Delta \omega$, since the extractable information to Bob will depend on how well he can distinguish the different basis states. For sub-unit efficiency, the basis states become mixed with the vacuum state, which diminishes their distinguishability, thus $\mathcal{C}$ must drop against loss. Note that for flat channel spectral response there is no bit-rate difference between the post-selected and non-post-selected cases. This is because the channel introduces no bias which enables the vacuum state to convey information about the encoded state. Thus, it makes no difference if it is post-selected away.

In the case of a Gaussian spectral response function, with perfect efficiency at \mbox{$\eta(0)=1$}, we observe that as the standard deviation of the spectral response function increases, so does $\mathcal{C}$. In the limit of large $\sigma$, the spectral response becomes flat with unity efficiency, \mbox{$\eta(\omega)\to 1$}, and with large $\Delta \omega$ we find the upper bound on the channel capacity is \mbox{$\mathcal{C}\leq\mathrm{log}_2 N$}, the theoretical maximum. In this limit Bob always observes the same state Alice transmitted, a basis of orthogonal pure states, and $\mathcal{C}$ saturates the bound when frequency-resolved photo-detection is employed by Bob.


For indistinguishable photons, no measurement performed by Bob is able to discern which encoded basis state is being transmitted, and thus the information capacity is zero for \mbox{$\Delta \omega=0$}. Similarly, for a very narrow channel spectral response function, Bob always measures the vacuum state and no information may be communicated. 

As expected, for Gaussian response, the post-selected channel bandwidth is strictly less than the non-post-selected bandwidth, owing to the information which is discarded during post-selection.

\begin{figure}[!htb]
\includegraphics[width=0.7\columnwidth]{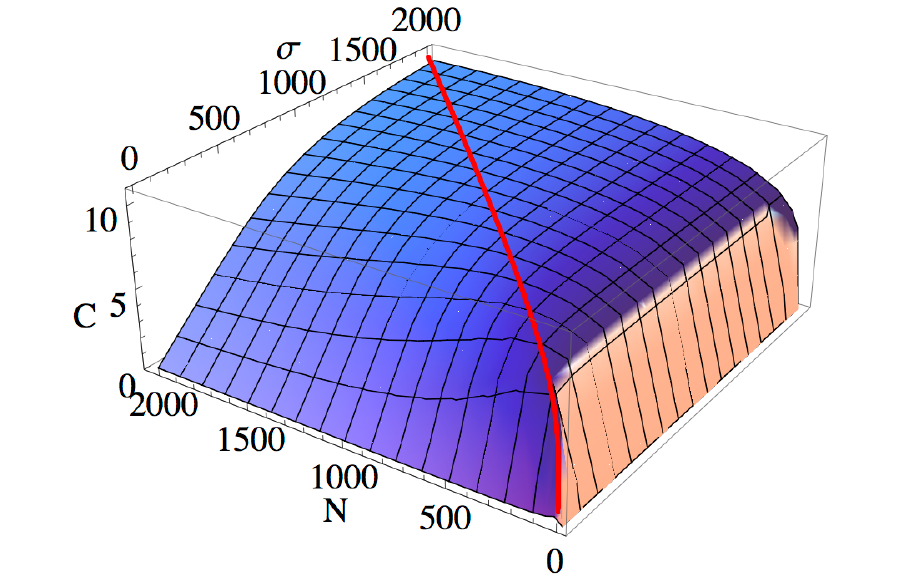}
\caption{(Colour online) Upper bound on the classical channel capacity (without post-selection) as a function of the alphabet size, $N$, and channel bandwidth, $\sigma$. The red line shows the optimal $N$ for a given $\sigma$. Note that the optimal encoding is calculated based on the Holevo bound, which only provides an upper bound on $\mathcal{C}$, which may not necessarily be saturated. Thus, the true optimal encoding may be different.} \label{fig:optimal}
\end{figure}

With a finite bandwidth channel there reaches a point where adding more basis states to the alphabet will not enhance the information capacity of the channel, since the additional letters reside in the region where \mbox{$\eta(\omega)\approx 0$}. On the contrary, it becomes counterproductive to employ additional letters since we are shifting the probability distribution within $\rho'$ into a region where no information may be communicated. In Fig. \ref{fig:optimal} we illustrate the relationship between the number of basis states, channel bandwidth, and $\mathcal{C}$. Evidently, for a given channel there is always a finite optimal value for $N$, shown by the red line. Thus, in general it is not optimal to always encode across the largest possible alphabet. Rather, the optimal alphabet size is a function of the channel spectral response.

\emph{Analytic bounds} --- We have previously considered the Holevo bound as a means for determining an upper bound on the channel capacity using a single photon. In certain circumstances, however, there is another option for bounding the capacity, which we now describe. Consider a photon encoded as \mbox{$\psi_i(\omega) \propto e^{-(d_i-\omega)^2/(4 \sigma_\psi^2)}$}, which passes through a channel with \mbox{$\eta(\omega) =  \sqrt{p_\text{peak}}e^{-\omega^2/(4 \sigma_\eta^2)}$}, a Gaussian scaled such that the peak transmission probability is $p_\text{peak}$.

The probability of the photon passing through the channel without being lost is $q_i = \int_{-\infty}^\infty \eta(\omega)^2 \psi_i(\omega)^2 \mathrm{d}\omega$, which in this case yields
\begin{equation}
q_i = \frac{p_\text{peak} }{\sigma_\psi \sqrt{\sigma_\psi^{-2} + \sigma_\eta^{-2}}} \exp\left(-\frac{d_i^2}{2(\sigma_\psi^2 + \sigma_\eta^2)}\right),
\end{equation}
and the state of the photon, if it is transmitted, has spectral distribution function $\psi_i'(\omega) = \frac{1}{\sqrt{q_i}} \eta(\omega) \psi_i(\omega)$.

Note that the minimum probability of photon loss obtained by maximising $q_i$ over $d$ for fixed $\sigma_\psi$ is 
\begin{equation}
q_\text{max} = \frac{p_\text{peak}}{\sigma_\psi \sqrt{\sigma_\psi^{-2} + \sigma_\eta^{-2}}},
\end{equation}
and hence an upper bound on the channel capacity is given by the channel capacity of a quantum erasure channel with erasure probability \mbox{$\kappa = 1-q_\text{max}$}. This channel has been studied extensively for the case of qubits, for which analytic results are known for both the quantum and classical capacities \cite{bennett1997capacities}. These arguments can trivially be extended to qudits to give the classical channel capacity ($\mathcal{C}$), quantum capacity ($\mathcal{Q}$), and quantum capacity assisted by two-way classical communication ($\mathcal{Q}_2$), all of which are equal to \mbox{$(1-\kappa)\log_2 N$} for the quantum erasure channel.

Thus, for the case of information encoded as a photon passing through a channel with a Gaussian transmission profile, where each letter is encoded as a Gaussian distribution over frequencies, all three capacities are bounded from above by
\begin{equation}
\mathcal{C},\mathcal{Q},\mathcal{Q}_2 \leq  \frac{p_\text{peak} \mathrm{log}_2 N}{\sigma_\psi \sqrt{\sigma_\psi^{-2} + \sigma_\eta^{-2}}}.
\end{equation}
In the alternate case of constant transmission probability, \mbox{$\eta(\omega) = \eta$}, as the overlap between the Gaussians encoding different letters tends to zero the channel approaches a quantum erasure channel and the capacities tend to \mbox{$\mathcal{C},\mathcal{Q},\mathcal{Q}_2 =  \eta^2 \mathrm{log}_2 N$} from below.

\begin{figure}[!htb]
\includegraphics[width=0.6\columnwidth]{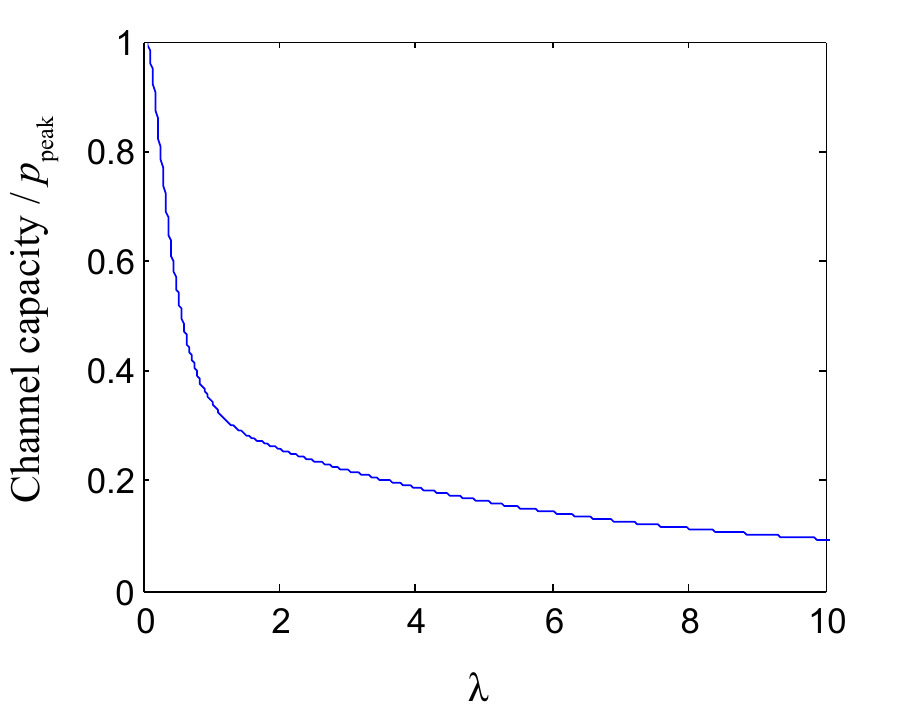}
\caption{Maximum classical channel capacity for two-state Gaussian encoding as a function of the ratio $\lambda = \sigma_\psi/\sigma_\eta$. \label{fig:2statecapacity}}
\end{figure}

Calculating exact channel capacities can be challenging. However, in the restricted case of two-state Gaussian encoding, it can be shown that the classical channel capacity is given exactly by,
\begin{equation}
\mathcal{C} = p_\text{peak} \frac{ e^{-\frac{\delta^2}{8(1+\lambda^2)}}}{\sqrt{1+\lambda^{2}}} I_2\left(\frac{1 - \sqrt{1 - e^{-\frac{\delta^2  (1 + \lambda^2)}{8 \lambda^2}}}}{2}\right),
\end{equation}
where \mbox{$\delta = \Delta\omega/\sigma_\eta$} and \mbox{$\lambda = \sigma_\psi/\sigma_\eta$}. Fig. \ref{fig:2statecapacity} shows the maximum value this capacity can take as a function of $\lambda$. The proof is given in the Supplementary Information.

\emph{Conclusion} --- We have discussed the scenario where a single photon, encoded in the spectral degree of freedom, communicates information between two parties using a channel with well-defined spectral response. We analytically derived upper bounds on the achievable classical and quantum channel capacities under a physically realistic model for the communications channel -- specifically, a channel with frequency-dependent absorption properties.

We noted that, in general, post-selection upon detection of a photon is sub-optimal, since measuring the vacuum state actually carries information about the encoded state sent by the transmitter. However, depending on the application, specifically in the context of a variable bit-rate channel, post-selection may be necessary. We calculated an upper bound on the classical channel capacity in the cases of both post-selected and non-post-selected communication, and on the quantum capacity in the case of Gaussian encoding.

We argued that, in general, it is not optimal to always encode across the largest possible alphabet. Rather, the optimal alphabet size will be a function of the channel spectral response.




\emph{Acknowledgments} --- We thank Gavin Brennen, Mark Byrd and Samuel Marks for helpful discussions. PR and AG acknowledge support from the Australian Research Council Centre of Excellence for Engineered Quantum Systems (Project number CE110001013). JF acknowledges support from the National Research Foundation and Ministry of Education, Singapore.


\bibliography{paper}


\appendix
\section{Supplementary information}

We now derive an analytic bound on the classical channel capacity in the restricted case of two-state encoding using Gaussian wave-packets. Consider a photon encoded as 
\begin{equation}
\psi_i(\omega) = \frac{1}{\sqrt{\sqrt{2 \pi} \sigma_\psi}}e^{-(d_i-\omega)^2/(4 \sigma_\psi^2)},
\end{equation}
which passes through a channel with 
\begin{equation}
\eta(\omega) =  \sqrt{p_\text{peak}}e^{-\omega^2/(4 \sigma_\eta^2)},
\end{equation}
a Gaussian scaled such that the peak transmission probability is $p_\text{peak}$.

The probability of the photon passing through the channel without being lost is 
\begin{eqnarray}
q_i &=& \int_{-\infty}^\infty \eta(\omega)^2 \psi_i(\omega)^2 \mathrm{d}\omega \nonumber \\
&=& \frac{p_\text{peak} }{\sigma_\psi \sqrt{\sigma_\psi^{-2} + \sigma_\eta^{-2}}} \mathrm{exp}\left(-\frac{d_i^2}{2(\sigma_\psi^2 + \sigma_\eta^2)}\right),
\end{eqnarray}
and the state of the photon, if it is transmitted, has spectral distribution function given by $\psi_i'(\omega) = \frac{1}{\sqrt{q_i}} \eta(\omega) \psi_i(\omega)$.

In general, the task of calculating the exact capacity of a communications channel is challenging, and has not been possible for most channels of interest. However, for the case of information encoded in one of two pure states, the classical capacity is known to be $I_2\left(\frac{1 - \sin(\alpha)}{2}\right)$ \cite{bennett1996parity}, where the overlap of the two states is $\cos(\alpha)$, and \mbox{$I_2(x) = 1 + x \log_2 x + (1-x) \log_2 (1-x)$}. Thus we have \mbox{$\mathcal{C} = I_2\Big(\frac{1 - \sqrt{1 - |\langle \psi_0 | \psi_1 \rangle |^2}}{2}\Big)$}.

If we consider a two letter alphabet encoded by two Gaussians of standard deviation $\sigma_\psi$ and separation between centres of $\Delta\omega$, then their overlap is $e^{-\Delta\omega^2/(8 \sigma_\psi^2)}$, and hence the classical capacity of the corresponding lossless channel will be given exactly by
\begin{equation}
\mathcal{C} = I_2\left(\frac{1 - \sqrt{1 - e^{-\Delta\omega^2/(4 \sigma_\psi^2)}}}{2}\right).
\end{equation}

If we consider the state after propagating through a channel $\eta(\omega)$ with a Gaussian transmission profile, as described earlier, then the output state for each encoded letter (if the photon is not absorbed) will be given by  $\psi_i'(\omega) = \frac{1}{\sqrt{q_i}} \eta(\omega) \psi_i(\omega)$. Since this is proportional to the product of two Gaussians, the result will be another Gaussian wave-packet with
\begin{equation}
\psi_i' (\omega) = \frac{1}{\sqrt{\sqrt{2 \pi} \sigma_{\psi\eta}}}e^{-(d_{i\eta}-\omega)^2/(4 \sigma_{\psi\eta}^2)},
\end{equation}
where \mbox{$\sigma_{\psi\eta} = \sqrt{\frac{\sigma_\psi^2 \sigma_\eta^2}{\sigma_\psi^2 + \sigma_\eta^2}}$} and \mbox{$d_{i\eta} = d_i \frac{\sigma_\eta^2}{\sigma_\psi^2 + \sigma_\eta^2}$}.

In order to ensure that the encoding works well in the post-selection regime, we ensure that detection or non-detection of a photon reveals no information about which letter is encoded by making the assumption that \mbox{$d_0 = -\Delta\omega/2$} and \mbox{$d_1 = \Delta\omega/2$}. The overlap between $\psi_0'$ and $\psi_1'$ is then given by \mbox{$e^{-\delta^2  (1 + \lambda^2)/(8 \lambda^2)}$}, where \mbox{$\delta = \Delta\omega/\sigma_\eta$} and \mbox{$\lambda = \sigma_\psi/\sigma_\eta$}. The classical capacity of this channel is thus
\begin{eqnarray}
\mathcal{C} &=& q_0  I_2\left(\frac{1 - \sqrt{1 - e^{-\frac{\delta^2  (1 + \lambda^2)}{8 \lambda^2}}}}{2}\right) \nonumber \\
&=& p_\text{peak} \frac{ e^{-\frac{\delta^2}{8(1+\lambda^2)}}}{\sqrt{1+\lambda^{2}}} I_2\left(\frac{1 - \sqrt{1 - e^{-\frac{\delta^2  (1 + \lambda^2)}{8 \lambda^2}}}}{2}\right).
\end{eqnarray}
Maximising this value over $\delta$ we obtain the exact maximum channel capacity as a function of $p_\text{peak}$ and $\lambda$.

\end{document}